\documentclass[aps,preprint]{revtex4}
\usepackage[dvips]{graphics,graphicx}
\begin{document}

\title{Hall conductivity beyond the linear response regime}
\author{Andrey R. Kolovsky$^{1,2}$}
\affiliation{$^1$ Kirensky Institute of Physics, 660036 Krasnoyarsk, Russia}
\affiliation{$^2$ Siberian Federal University, 660036 Krasnoyarsk, Russia}
\date{\today}

\begin{abstract}
The paper introduces a semi-analytical method for calculating the Hall conductivity in the single-band approximations. The method goes beyond the linear response theory and, thus, imposes no limitation on the electric fields magnitude. It is shown that the Hall current decreases with increase of the electric field, if the Bloch frequency (which is proportional to the electric field) exceeds the cyclotron frequency (which is proportional to the magnetic field). The obtained results can be directly applied to the system of cold Bose atoms in a 2D optical lattice, where the single-band approximation is well justified.
\end{abstract}
\maketitle

\section{Introduction}

Begging from works by Ohm in XIX century \cite{Ohm1927} and till seventeenth of XX century all studies of ordinary conductivity in solid crystals were restricted by the weak field regime, where the electric field is considered as a perturbation. This approach was actually justified because for typical laboratory conditions the Bloch frequency, which is proportional to the electric field, is much smaller than the characteristic relaxation rate in a crystal. The situation has changed in 1970, when Esaki and Tsu reported first measurements of the Ohm current in semiconductor superlattices \cite{Esak70}.  It was found that with increase of the electric field the current reaches some maximum value and then decreases -- the phenomenon known nowadays as the negative differential conductivity. This pioneering work initiated experimental and theoretical studies of ordinary conductivity in the strong field regime, where the electric field cannot be considered as a perturbation. We note that, besides semiconductor superlattices, the negative differential conductivity was also observed for cold neutral atoms in (quasi) 1D optical lattices subjected to a static force \cite{Ott04}. A great advantage of the latter system over semiconductor superlattices is full experimental control over relaxation processes. Because of this control one can study with cold atoms both the Hamiltonian and dissipative dynamics of the carriers, i.e, Bloch oscillations and Ohmic current.  

A different yet related transport problem is the Hall current in quantum dot and antidot arrays (see, \cite{Gerh89,Flei92,Weis94,Ishi95}, to cite a few of hundreds relevant papers).  These systems can be considered as 2D superlattices, where the 2D electron gas is subjected to a periodic potential with period of few hundreds nanometers. Here researches mainly focus on the effect of a magnetic field rather than on Bloch oscillations \cite{suris}. Indeed, because of a large superlattice period already a modest magnetic field $\sim1$T causes dramatic changes in the electron energy spectrum. This new spectrum is then substituted in the Kubo-type equation to calculate the conductivity tensor $\sigma_{ij}=\sigma_{ij}(B)$. During the last two decades  many exciting theoretical and experimental results on the Hall conductivity in quantum-dot arrays were reported. These studies revealed very nontrivial dependence of the conductivity tensor on the magnetic field and the Fermi energy. Let us also note that in a typical quantum-dot array the Fermi energy of the 2D electron gas is in a Bloch (mini)band with large index $n>10$, so that the single-band approximation, which is known to result in the famous Hofstadter's butterfly spectrum \cite{Hofs76}, is not applicable. Because of this complication of the problem almost all theoretical studies on the Hall conductivity in quantum-dot arrays have been restricted by the linear response regime \cite{remark0}.


Recently it has been noticed in Ref.~ \cite{Jaks03} that the Hofstadter butterfly can be realized with cold atoms in a 2D optical lattice by introducing an artificial magnetic field, which imposes the Lorentz force on moving neutral atoms.  To create this artificial field the authors of the cited paper suggested to use Raman transitions between the internal atomic states. This idea of an artificial magnetic field was developed further in other publications and today there are several different theoretical proposals for creating effective magnetic fields for atoms in a lattice (see, for example, the recent paper \cite{preprint} and references therein) and a successful realization of one of schemes for atoms in a harmonic trap \cite{Lin09}. Thus this is a matter of time when experimentalists will be able to mimic magnetic fields for cold atoms in a lattice, in the same manner as they now routinely mimic electric fields. Since optical lattices are much cleaner and controllable systems than quantum-dot arrays, this will open wide perspectives for studying the Hall conductivity in different parameter regimes, including the nonlinear response regime. The present work precedes these anticipated experimental studies. It presents a semi-analytical method for calculating the Hall current in the single-band approximation which, as mentioned above, can be easily justified for cold atoms in optical lattices. The method treats magnetic and electric fields on equal footing and, hence, goes beyond the linear response theory. In this sense we extend the Esaki-Tsu approach to the Ohm current in 1D lattices onto the Hall current in 2D lattices.

\section{Esaki-Tsu dependence for the Ohm current}

First we recall the reader few results on ordinary conductivity in the non-perturbative regime. To explain the negative differential conductactivity in semiconductor superlattices Esaki and Tsu used a kind of semiclassical approach, which resulted in the following dependence for the Ohm current:
\begin{equation} 
\label{0}
\frac{v}{v_0}= f(T) \frac{\omega_B/\gamma}{1+(\omega_B/\gamma)^2} \;, \quad
\omega_B=\frac{edF}{\hbar} \;.
\end{equation}
In Eq.~(\ref{0}) $F$ is the electric field, $d$ the lattice period, $e$ the charge, $\gamma$ the relaxation constant, and the pre-factor $f(T)$ accounts for the temperature dependence of the current [$f(0)=1$]. A microscopic derivation of the Esaki-Tsu  dependence (\ref{0}) was given by Minot in 2004 \cite{Mino04}. This was obtained by solving the master equation for the carriers one-particle  density matrix $\hat{\rho}$:
\begin{equation} 
\label{0a}
\frac{d\hat{\rho}}{dt}=-\frac{i}{\hbar}[\widehat{H},\hat{\rho}] -\gamma(\hat{\rho} - \hat{\rho}_0) \;.
\end{equation}
In this equation $\widehat{H}$ is the single-particle Hamiltonian of a carrier in a biased superlattice,
\begin{equation} 
\label{0b}
\widehat{H}=\widehat{H}_0 + edF\sum_{l} |l\rangle l \langle l| \;,\quad 
\widehat{H}_0=-\frac{J}{2}\sum_{l} \left( |l+1\rangle\langle l| +h.c.\right)
\end{equation}
and $\hat{\rho}_0$ is the equilibrium density matrix for $F=0$,
\begin{equation}
\label{0c}
\hat{\rho}_0 \sim \exp(-\beta\widehat{H}_0) \;,\quad \beta=1/k_B T \;.
\end{equation}
The model (\ref{0a}-\ref{0c}) results in Eq.~(\ref{0}) with the correct pre-factor  $f(T)={\cal I}_1(\beta J)/{\cal I}_0(\beta J)$.

The above microscopic derivation of the Esaki-Tsu dependence was revisited in Ref.~\cite{72} with respect to the problem of ordinary conductivity with cold atoms in 1D optical lattices. A weak point of the master equation (\ref{0a}) is that it is not in the Lindblad form. (Exclusions are the cases of zero and infinite temperature, where it can be rewritten in the Lindblad form.) Because of this drawback it may give wrong result for the velocity distribution of the carriers.  However, it was confirmed that it gives qualitatively correct result for the mean velocity, i.e., the current.  The master equation (\ref{0a}) will be our theoretical framework in studying the Hall conductivity in 2D lattices.

\section{The model}
 
We consider a quantum particle in a square lattice of side $d$ in the $x-y$ plane. The particle is subjected to an in-plane electric field $F$, aligned with the $y$ axis, and a magnetic field $B$ normal to the $x-y$ plane. Using the tight-binding approximation and the gauge ${\bf A}=B(-y,0,0)$ for the vector potential the particle Hamiltonian reads
\begin{equation}
\label{1a}
\widehat{H}= \widehat{H}_0 +edF\sum_{l,m} |l,m\rangle m \langle l,m |  \;,
\end{equation}
\begin{equation}
\label{1b}
\widehat{H}_0= -\frac{J_x}{2} \sum_{l,m} \left(|l+1,m\rangle \langle l,m | e^{i2\pi\alpha m} + h.c.\right)
-\frac{J_y}{2} \sum_{l,m} \left(|l,m+1\rangle \langle l,m |  + h.c.\right)  \;.
\end{equation}
The dimensionless parameter $\alpha$ in (\ref{1b}) is the Peierls phase, which is given by number of magnetic flux quanta per unit-cell area, $\alpha=eBd^2/hc$.  Besides the Peierls phase $\alpha$ and the Bloch frequency $\omega_B=edF/\hbar$ the other important characteristics of the system are the carrier effective mass, $M^*=d^2(J_xJ_y)^{1/2}/\hbar^2$, the cyclotron frequency, $\omega_c=eB/cM^*=2\pi \alpha (J_xJ_y)^{1/2}/\hbar$, and the drift velocity $v^*=ceF/B=d^2 eF/h\alpha$.  We note that for a charged particle (electron in a solid crystal) the Hamiltonian (\ref{1b}) is justified only in the limit of small $\alpha$, where the cyclotron radius of the classical orbit essentially exceeds the lattice period. This is, however, not the case for cold atoms in optical lattices, where the actual parameter of the system is the Peierls phase but not the magnitude of a magnetic field \cite{Jaks03,preprint}. Hence, we impose no limitations on $\alpha$ and, without any loss of generality, one may  consider  $|\alpha|\le 1/2$.

Our aim is to calculate the Hall  ($v_x$) and the Ohm ($v_y$) currents,
\begin{equation} 
\label{2a}
v_{x,y}={\rm Tr}[\hat{v}_{x,y}\hat{\rho}_{st}] \;,
\end{equation}
where $\hat{\rho}_{st}$ is the stationary solution of the master equation (\ref{0a}) and $\hat{v}_{x,y}$ the current operators,
\begin{equation}
\label{2aa}
\hat{v}_x=-\frac{i}{\hbar}[\widehat{H}_0,\hat{x}] \;,\quad \hat{x}=d\sum_{l,m} |l,m\rangle l \langle l,m| \;,
\end{equation}
and for $\hat{v}_y$ one has a similar expression. Substituting  (\ref{1b}) in (\ref{2aa}) we have
\begin{equation}
\label{2b}
\hat{v}_x=\frac{v_0^{(x)}}{2i}\sum_{l,m}\left(|l+1,m\rangle\langle l,m| e^{i2\pi\alpha m} -h.c.\right) \;,\quad
v_0^{(x)}=\frac{d J_x}{\hbar} \;, 
\end{equation}
and
\begin{equation}
\label{2bb}
\hat{v}_y=\frac{v_0^{(y)}}{2i}\sum_{l,m} \left( |l,m+1\rangle\langle l,m| -h.c.\right) \;,
\quad v_0^{(y)}=\frac{d J_y}{\hbar} \;.
\end{equation}
%

\section{Landau-Stark states}

We shall perform calculations in the basis of the Landau-Stark states, which are the eigenstates of the Hamiltonian (\ref{1a}). To simplify equations, from now on we set the lattice period $d$ and the Planck constant $\hbar$ to unity. 

One finds the Landau-Stark states semi-analytically by using the following ansatz \cite{Muno05}:
\begin{equation}
\label{3a}
\psi_{l,m}= \frac{e^{i\kappa l}}{\sqrt{L_x}} b_m   \;.
\end{equation}
In Eq.~(\ref{3a}) $\kappa=2\pi k/L_x$ is the quasimomentum, $0\le \kappa<2\pi$, $L_x$ is the lattice size in the $x$ direction and we eventually let $L_x$ tend to infinity.  Substituting (\ref{3a}) into the stationary Schr\"odinger equation with the Hamiltonian (\ref{1a}),  
\begin{equation}
\label{3b}
-\frac{J_x}{2}\left(e^{-i2\pi\alpha m} \psi_{l+1,m}  +  e^{i2\pi\alpha m} \psi_{l-1,m}\right)
-\frac{J_y}{2}\left(\psi_{l,m+1} + \psi_{l,m-1}\right)
+Fm \psi_{l,m}=E\psi_{l,m} \;,
\end{equation}
we reduce it to the following 1D eiqenvalue problem:
\begin{equation}
\label{3c}
-\frac{J_y}{2}(b_{m+1}+b_{m-1}) + [Fm - J_x\cos(2\pi\alpha m-\kappa)]b_m =E b_m \;.
\end{equation}
Equation (\ref{3c}) is a kind of the 1D Wannier-Stark problem and can be easily solved numerically. Labeling the solution by the discrete index $\nu$ and scanning over the quasimomentum $\kappa$ we find the energy spectrum $E=E_\nu(\kappa)$ and the Landau-Stark states $|\Psi_{\nu,\kappa}\rangle=\sum_{l,m} \psi_{l,m}^{(\nu,\kappa)} |l,m\rangle$.

\begin{figure}
\center
\includegraphics[width=12cm,clip]{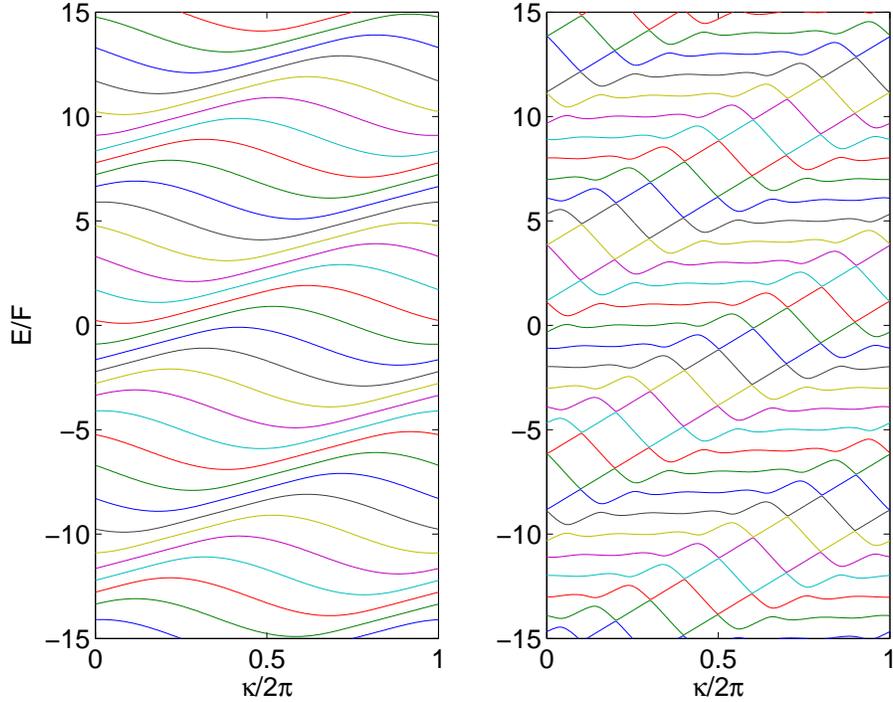}
\caption{A fragment of the energy spectrum of the Landau-Stark states for $J_x=J_y=1$, $\alpha=1/10$, and $F=1$ (left) and $F=0.3$ (right). The figure is borrowed from Ref.~\cite{preprint2}.}
\label{fig1}
\end{figure}

The energy spectrum and properties of the Landau-Stark states were studied in some detail in our recent work \cite{preprint2} devoted to the Hamiltonian dynamics of the system (\ref{1a}). 
As an example, Fig.~\ref{fig1} shows the energy spectrum  $E=E_\nu(\kappa)$ for $J_x=J_y=1$, $\alpha=1/10$, and two different values of $F$. This figure is aimed to illustrate a qualitative change in the spectrum, which takes place around
\begin{equation}
\label{33}
F_{cr}=2\pi\alpha J_x  \;.
\end{equation}
Namely, for $F<F_{cr}$ the energy form a pattern with straight lines. The Landau-Stark states belonging to these lines are the transporting states, which transport the quantum particle in orthogonal to the field direction with the drift velocity $v^*$,
\begin{equation}
\label{34}
v^*=F/2\pi\alpha  \;.
\end{equation}
%

Having the Landau-Stark states obtained we calculate the current operators (\ref{2b}) and (\ref{2bb})  in this basis. We have
\begin{equation}
\label{4a}
\langle \Psi_{\nu,\kappa}|\hat{v}_x|\Psi_{\nu',\kappa'}\rangle
= \delta(\kappa-\kappa') J_x 
\sum_m b_m^{(\nu)}(\kappa)b_m^{(\nu')}(\kappa) \sin(2\pi\alpha m -\kappa) 
\equiv  \delta(\kappa-\kappa') V_{\nu,\nu'}^{(x)}(\kappa) \;,  
\end{equation}
and
\begin{equation}
\label{4aa}
\langle \Psi_{\nu,\kappa}|\hat{v}_y|\Psi_{\nu',\kappa'}\rangle
= \delta(\kappa-\kappa') \frac{J_y}{2i}
\sum_m \left[b_{m+1}^{(\nu)}(\kappa)- b_{m-1}^{(\nu)}(\kappa)\right] b_m^{(\nu')}(\kappa) 
\equiv  \delta(\kappa-\kappa') V_{\nu,\nu'}^{(y)}(\kappa) \;.
\end{equation}
Because of the presence the $\delta$-function in (\ref{4a},\ref{4aa}), Eq.~(\ref{2a}) for the Hall current simplifies as  
\begin{equation} 
\label{4b}
v_{x,y}=\frac{1}{2\pi}\int_0^{2\pi} {\rm d}\kappa {\rm Tr}[V^{(x,y)}(\kappa){\cal R}^{(st)}(\kappa)] \;,
\end{equation}
where ${\cal R}^{(st)}(\kappa)$ is the $\kappa$-specific stationary density matrix,
\begin{equation}
\label{4c}
{\cal R}^{(st)}_{\nu,\nu'}(\kappa)=\langle \Psi_{\nu,\kappa}|\hat{\rho}_{st}|\Psi_{\nu',\kappa}\rangle
=\frac{\gamma}{\gamma+i[E_{\nu'}(\kappa)-E_\nu(\kappa)]} {\cal R}^{(0)}_{\nu,\nu'}(\kappa) \;.
\end{equation}

\section{Landau states}

Next we specify the equilibrium density matrix $\hat{\rho}_0$. To have tractable results we shall consider the case where only the lowest Landau levels are populated. Thus we assume
\begin{equation}
\label{5}
\hat{\rho}_0=\frac{1}{{\cal N}}\sum_{j=1}^{\cal N} |\Phi_j\rangle\langle \Phi_j| \;,
\end{equation}
where ${\cal N}=L_y L_x \alpha$ and $|\Phi_j\rangle$ are the lowest energy Landau states. The density matrix (\ref{5}) corresponds to ${\cal N}$ fermionic carriers at zero temperature. Alternative, it may be considered as a density matrix of non-interacting bosons at a finite temperature. In what follows we adopt the latter point of view, where the relevant temperature interval is discussed in the last paragraph of the section. We would like to stress that our choice of the equilibrium density matrix is exclusively for the sake of easy interpretation of numerical results. In principle, one can consider an arbitrary $\hat{\rho}_0$.  This way the reported below results can be generalized to arbitrary temperature and arbitrary carrier statistics. 
\begin{figure}[t]
\center
\includegraphics[width=8cm,clip]{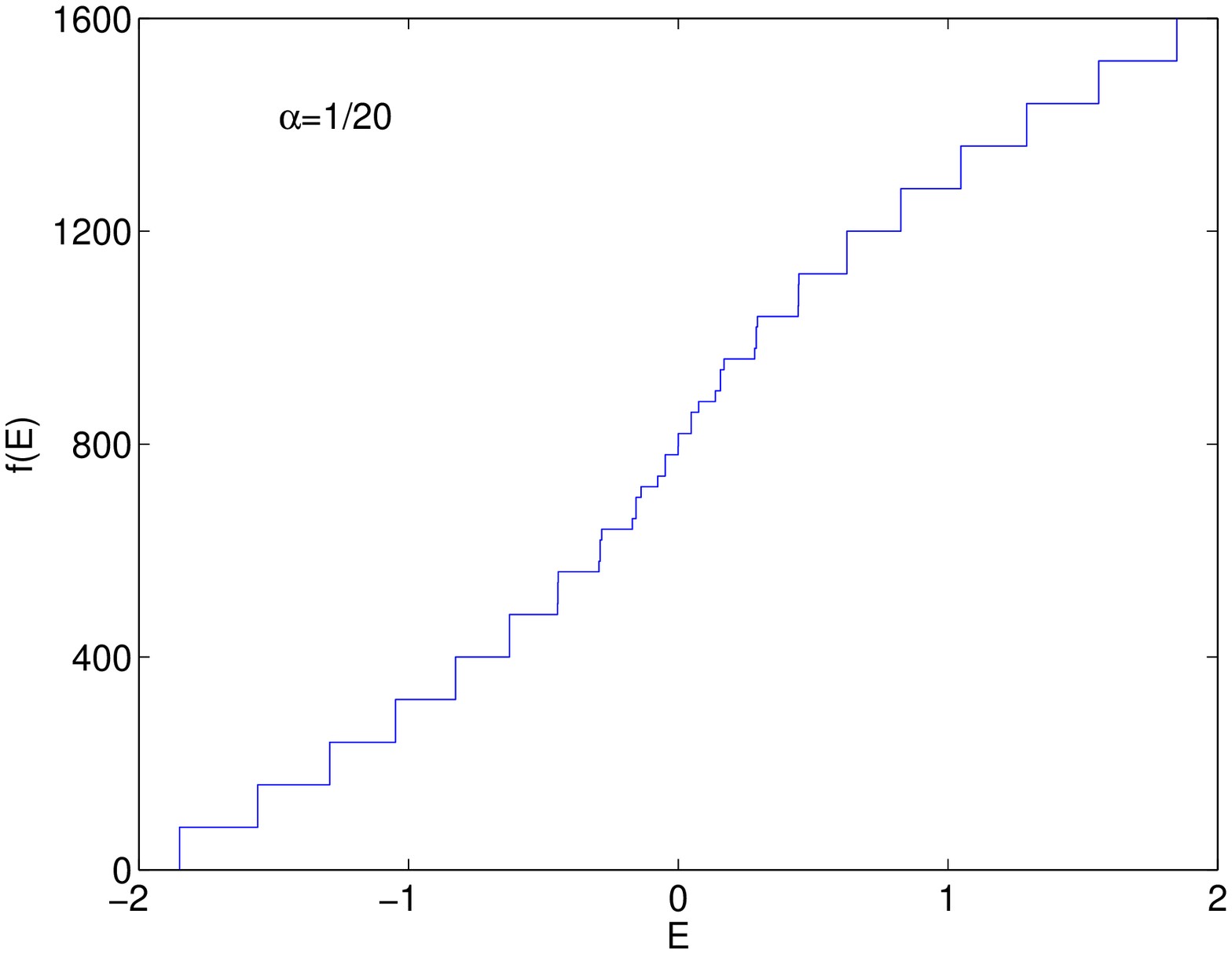}
\includegraphics[width=8cm,clip]{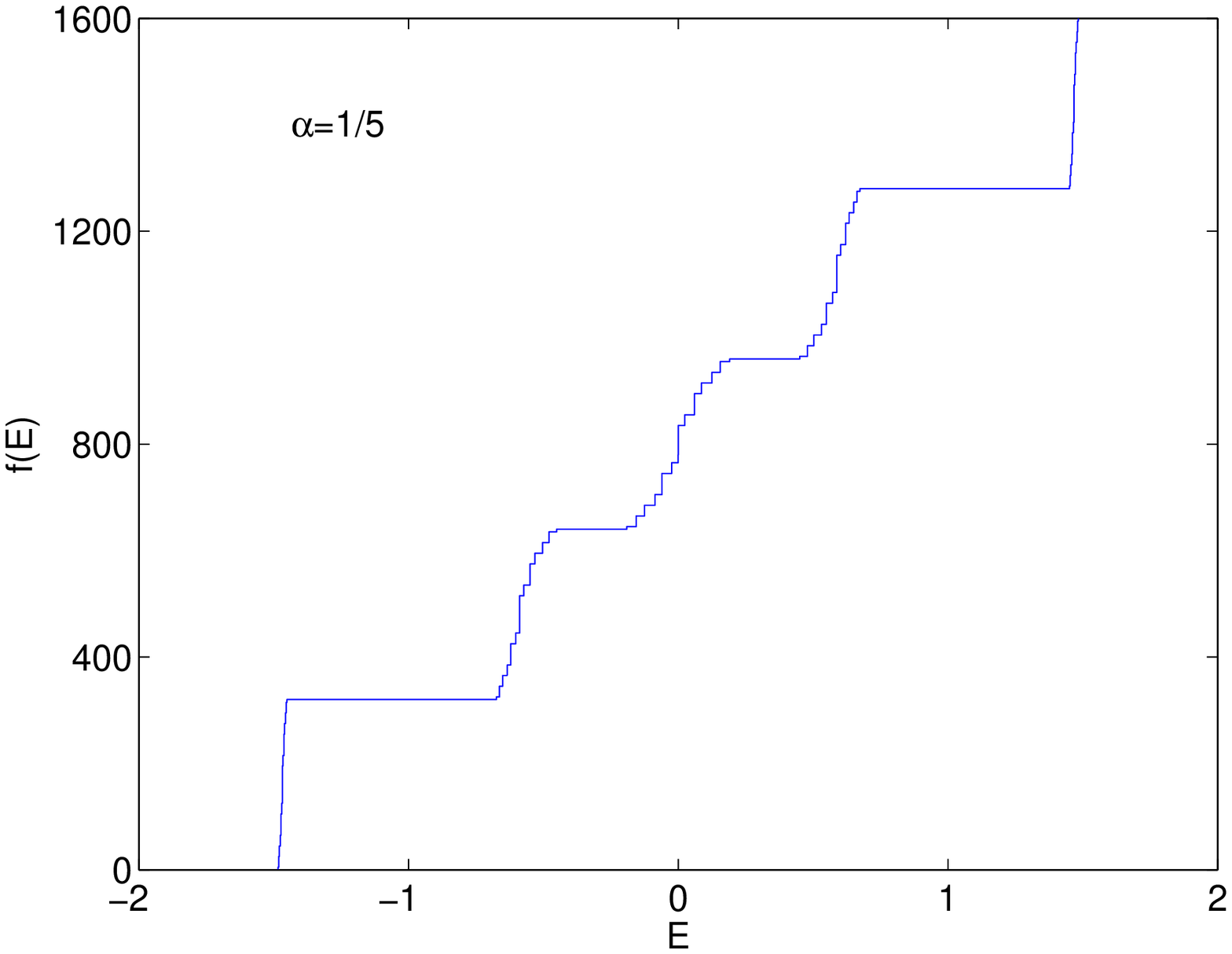}
\caption{Integrated density of states of the system (\ref{1b}) for $\alpha=1/20$ (left) and $\alpha=1/5$ (right). The other parameters are $J_x=J_y=1$ and $L_x=L_y=40$.}
\label{fig2}
\end{figure}

Similar to the case of Landau-Stark states, one finds the Landau states semi-analytically by using the substitution (\ref{3a}), which reduces the eigenvalue problem for the Hamiltonian $\widehat{H}_0$ to that for Harper's Hamiltonian,
\begin{equation}
\label{6b}
-\frac{J_y}{2}(b_{m+1}+b_{m-1}) -J_x\cos(2\pi\alpha m-\kappa) b_m  = E b_m \;.
\end{equation}
Figure \ref{fig2} shows the integrated density of states of the system (\ref{1b}) calculated on the basis of Eq.~(\ref{6b}) for $\alpha=1/20$ and $\alpha=1/5$.  The states $|\Phi_j\rangle$ in (\ref{5}) are associated with the first step in the integrated density. It is also easy to show that the length of this step is approximately given by the cyclotron energy $\hbar\omega_c=2\pi\alpha(J_xJ_y)^{1/2}$. 
Thus our condition on the temperature reads  $k_B T \ll \hbar\omega_c$. At the same time, to have equal populations of the lowest Landau states, we assume $k_BT \gg \Delta$, where $\Delta$ is the width of the lowest magnetic band \cite{remark1}. As seen in Fig.~\ref{fig2} the required condition $\Delta \ll k_BT \ll \hbar\omega_c$ is easier to satisfy for small $\alpha$.

\section{Numerical procedure and results}

The numerical procedure is as follows.  We fix  $L_x$ and $L_y$ and calculate the Landau and Landau-Stark states. The lattice size $L_x$ defines the discrete step for the quasimomentum, which should be small enough to resolve main quasi-crossings in the energy spectrum in Fig.~\ref{fig1}. The lattice size $L_y$ is arbitrary yet, to reduce the boundary effect when solving (\ref{3c}), $L_y\gg 1/\alpha$. Next we calculate the $\kappa$-specific matrices of the current operators and the stationary density matrix (\ref{4c}). We note that for a rational $\alpha=r/q$ the infinite matrix of the current operators  as well as $\kappa$-specific density matrix obey the translational symmetry,
\begin{equation}
\label{7}
V^{(x,y)}_{\nu'+q,\nu+q}(\kappa)=V^{(x,y)}_{\nu',\nu}(\kappa) \;, \quad
{\cal R}^{(st)}_{\nu'+q,\nu+q}(\kappa)={\cal R}^{(st)}_{\nu',\nu}(\kappa) \;,
\end{equation}
which further facilitates the numerical procedure. Finally, substituting these matrices into (\ref{4b}) and integrating over the quasimomentum $\kappa$ we calculate the Hall and Ohm currents. 
\begin{figure}
\center
\includegraphics[width=8cm,clip]{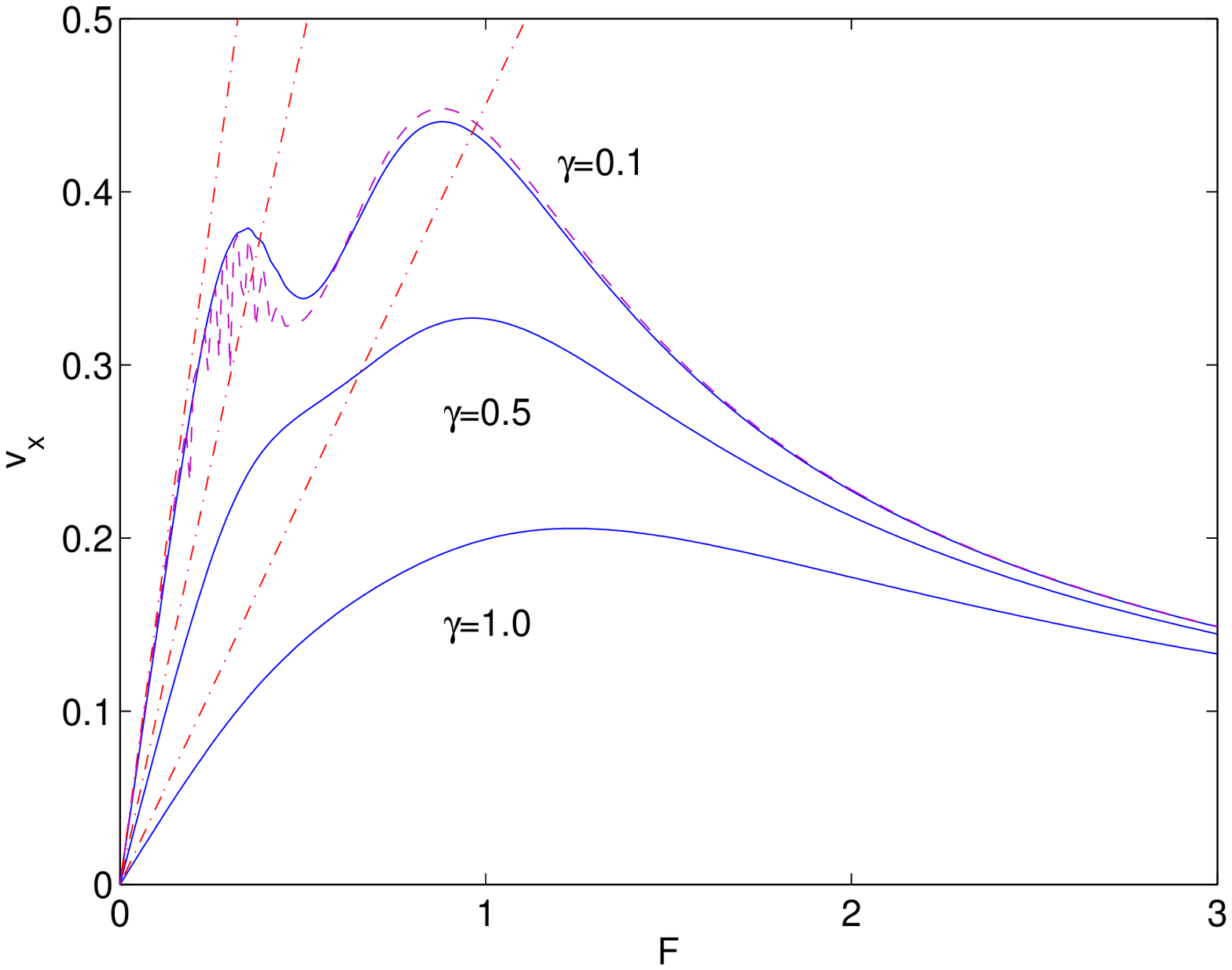}
\includegraphics[width=8cm,clip]{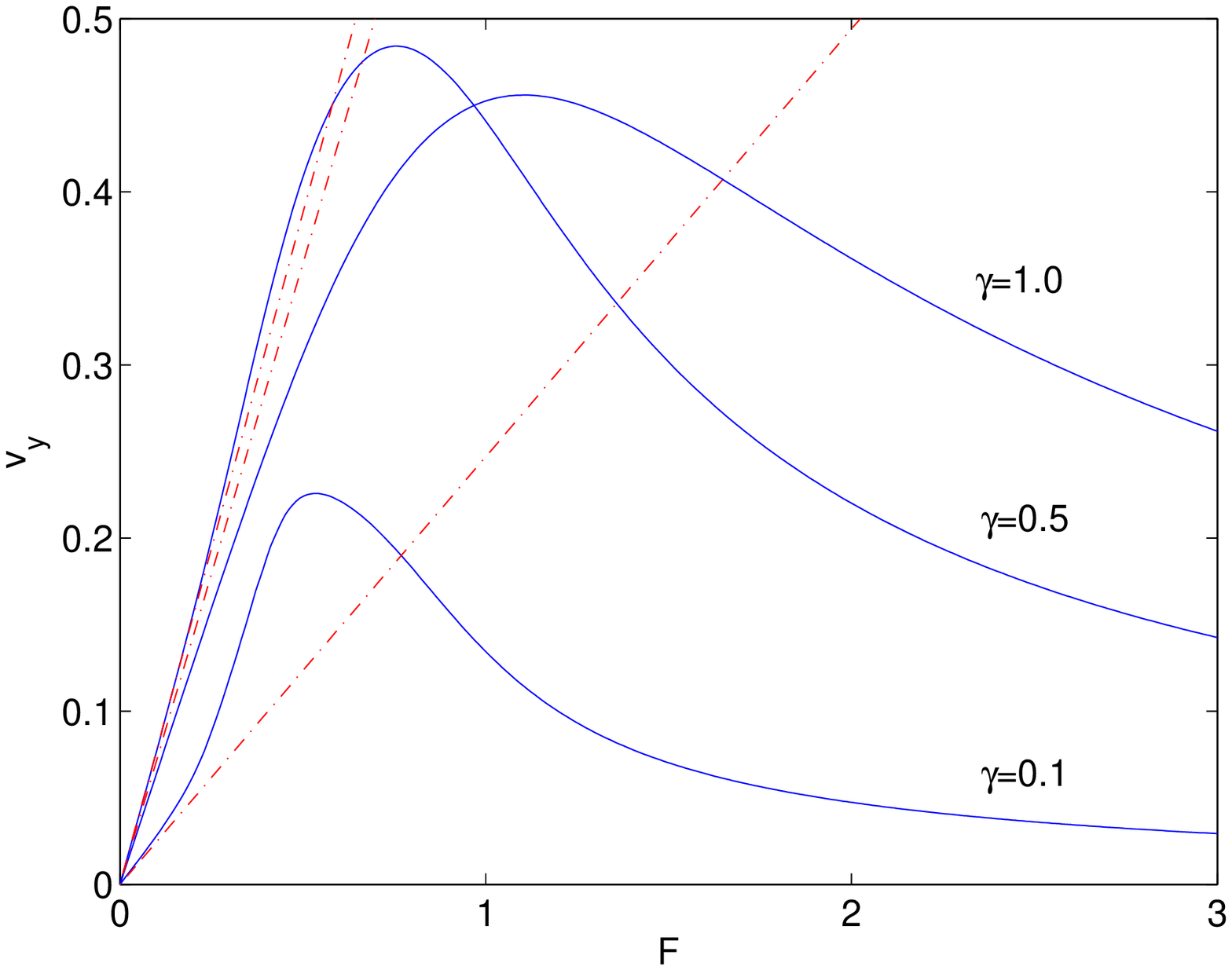}
\caption{The Hall (left) and Ohm (right) current as the function of electric field $F$ for different relaxation rates $\gamma$. The other parameters are $J_x=J_y=1$ and $\alpha=1/10$, the lattice size $L_x=L_y=40$. The straight dash-dotted lines are predictions of the linear response theory. Additional dashed line in the left panel shows the Hall current for $\gamma=0$, where the Ohm current vanishes. }
\label{fig3}
\end{figure}

\begin{figure}
\center
\includegraphics[width=8cm,clip]{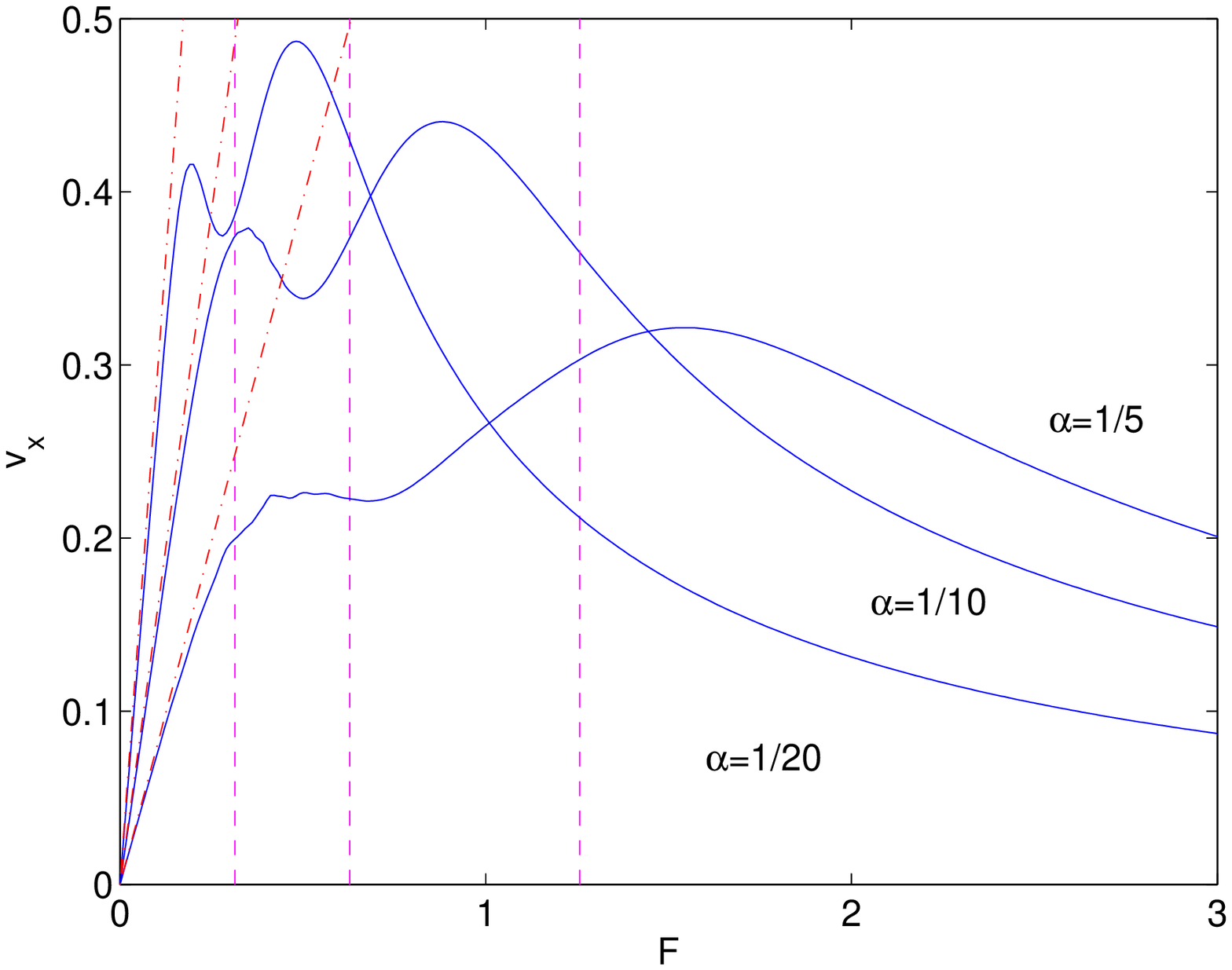}
\includegraphics[width=8cm,clip]{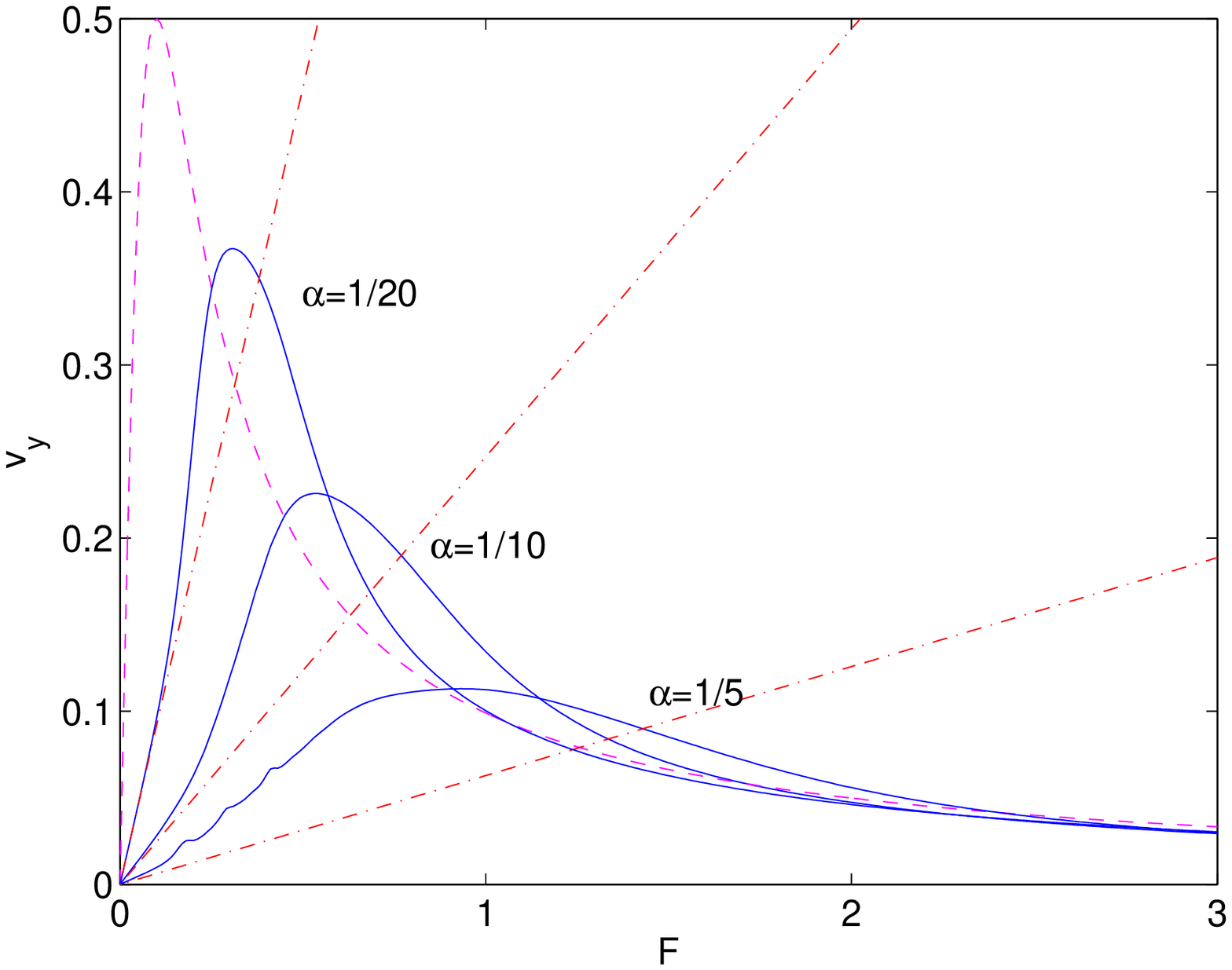}
\caption{The Hall (left) and Ohm (right) current as the function of electric field $F$ for $\gamma=0.1$ and different $\alpha$.  Additional dashed line in the right panel shows the Ohm current for $\alpha=0$, where the Hall current vanishes.}
\label{fig4}
\end{figure}

The left panel in Fig.\ref{fig3} shows the Hall current $v_x$ as the function of the applied field $F$ for $\alpha=1/10$ and different values of the relaxation constant $\gamma$. We begin with considering the case $\gamma=0$ (dashed line), which corresponds to the Hamiltonian dynamics of the carriers. As shown in Ref.~\cite{preprint2}, for the specified initial conditions (population of the ground Landau states) a weak static field transports the carriers in the orthogonal direction with the drift velocity (\ref{34}). Thus in the weak field regime the dependence $v_x=v_x(F)$ is approximated by $v_x=F/2\pi\alpha$. The increase in the Hall current continues till $F$ reaches some critical value $F^*$, where the function $v_x=v_x(F)$ has the global maximum. An estimate for $F^*$ is provided by Eq.~(\ref{33}), although we found $F^*$ to be systematically larger than $F_{cr}$ by a numerical factor $1<z<2$ [see Fig.~\ref{fig4}(a) below]. With further growth of the electric field we enter the regime of negative differential conductivity, where the Hall current decreases with increase of $F$.

As mentioned above, one finds an explanation for the transition from positive to negative differential conductivity regimes in structural changes of the Landau-Stark states, which take place around $F_{cr}$ \cite{preprint2}. We also give another explanation, which is based on the Landau states picture. Namely, using the Kramers-Hennenberger transformation an electric field is seen as periodic driving of the system with the Bloch frequency $\omega_B=F$. When $\omega_B$ matches the energy gap between the ground and the first magnetic band [i.e., the length of the first step in Fig.~\ref{fig2}, approximately given by the cyclotron frequency $\omega_c$], the driving induces transitions between the Landau levels and we observe the local minimum in the dependence $v_x=v_x(F)$ which always precedes the global maximum. Thus the necessary condition for the negative differential conductivity regime can be also formulated as a requirement that the Bloch frequency exceeds the cyclotron frequency. 

The other (solid) lines in Fig.~\ref{fig3} show the dissipative Hall current for $\gamma=0.1,0.5,1$. It is seen that a finite relaxation rate suppresses the Hall current and smoothes fine features of the dependence $v_x=v_x(F)$ for the non-dissipative Hall current. In addition to Fig.~\ref{fig3}(a) figure \ref{fig4}(a) shows the Hall current for the fixed $\gamma=0.1$ and different $\alpha$. The vertical dashed lines in this figure indicate the critical electric field (\ref{33}) for each case. 

The right panels in Fig.~\ref{fig3} and Fig.~\ref{fig4} show the Ohm current. It is seen in Fig.~\ref{fig3} that larger relaxation rates suppress the Hall current but enhance the Ohm current. In the limit $\gamma\rightarrow\infty$ the Hall current vanishes and the dependence $v_y=v_y(F)$ for the Ohm current approaches the Esaki-Tsu dependence (\ref{0}). Alternatively, one recovers the Esaki-Tsu result by considering the limit $\alpha\rightarrow0$, see Fig.~\ref{fig4}(b). In Fig.~\ref{fig3} and Fig.~\ref{fig4} we also depict predictions of the linear response theory, ${\bf v}=\sigma {\bf F}$, where the off-diagonal and diagonal elements of the conductivity tensor  are given by the Drude-type formulas,
\begin{equation}
\label{9}
\sigma_{xy}=\frac{1}{\gamma}\frac{\omega_c/\gamma}{1+(\omega_c/\gamma)^2} \;,\quad
\sigma_{yy}=\frac{1}{\gamma}\frac{1}{1+(\omega_c/\gamma)^2} \;,
\end{equation}
and we approximate the cyclotron frequency by  $\omega_c=2\pi\alpha(J_x J_y)^{1/2}$.

\section{Conclusions}

We considered the quantum particle in a 2D lattice subjected to (real or artificial) electric and magnetic fields and calculated the Hall and Ohm currents as functions of the electric field magnitude. Although the obtained dependence $v_x=v_x(F)$ for the Hall current resembles the Esaki-Tsu dependence $v_y=v_y(F)$ for the Ohm current in the absence of magnetic field, the physics behind these two dependences is completely different. Indeed, the Esaki-Tsu dependence for the Ohm current appears due to an interplay between Bloch oscillations and relaxation processes and the Ohm current vanishes if $\gamma=0$. Conversely, the Hall current in the transverse direction takes place even in the absence of dissipation. The actual reason for the Esaki-Tsu like dependence for the Hall current is a qualitative change in the structure of the Landau-Stark states, which happens around $F_{cr}$. Note that the condition $F=F_{cr}$ means the Bloch frequency to coincide with the cyclotron frequency. Thus the Ezaki-Tsu like dependence for the Hall current is the result of an interplay between Bloch and cyclotron oscillations but not Bloch oscillations and relaxation processes. 

Concluding the discussion we would like to stress that in this work we do not addresss the quantum Hall effect. The latter phenomenon occurs for fermionic carriers when the magnetic field or the Fermi energy are varied. It would be interesting to study the quantum Hall effect in the non-perturbative regime, where the conductivity tensor depends on the electric field magnitude.  This problem has been addressed in already cited papers \cite{Kuno03,Kuno09}. However, these papers analyze the Hall conductivity specifically with respect to semiconductor structures and employ a different model.  

\vspace*{5mm}
\noindent
{\it Acknowledgments}\\
This work was supported by Russian Foundation for Basic Research, grant RFBR-10-02-00171-a, and by Deutsche Forschungsgemeinschaft via the Graduiertenkolleg `Nichtlineare Optik und Ultrakurzzeitphysik'.



\begin{thebibliography}{0}
\bibitem{remark0}
An example of rare exclusions are works \cite{Kuno03,Kuno09}, which are briefly discussed in the concluding section.
\bibitem{remark1}
Here we use the notion of the magnetic band irrespectively to the commensurability condition. In the other words, $\Delta$ is the width of the vertical part of a step in the integrated density of state, which is essentially the same for a rational $\alpha$ and irrational $\alpha'\approx\alpha$.
\end{thebibliography}
\end{document}